\documentclass[letterpaper,english,reprint,superscriptaddress,longbibliography]{revtex4-1}
\usepackage{lmodern}

\usepackage[T1]{fontenc}
\usepackage[utf8]{inputenc}
\usepackage{babel}
\usepackage{amsmath}
\usepackage{amssymb}
\usepackage{graphicx}
\usepackage[pdfusetitle,
 bookmarks=false,
 breaklinks=false,pdfborder={0 0 0},pdfborderstyle={},backref=false,colorlinks=false]
 {hyperref}

\makeatletter

\pdfpageheight\paperheight
\pdfpagewidth\paperwidth

\makeatother

\begin{document}
\title{Imaging transverse modes in a GHz surface acoustic wave cavity}
\author{M. Fisicaro}
\email{fisicaro@physics.leidenuniv.nl}

\affiliation{Huygens-Kamerlingh Onnes Laboratory, Leiden University, P.O. Box 9504,
2300 RA Leiden, The Netherlands}
\author{T. A. Steenbergen}
\affiliation{Huygens-Kamerlingh Onnes Laboratory, Leiden University, P.O. Box 9504,
2300 RA Leiden, The Netherlands}
\author{Y. C. Doedes}
\affiliation{Huygens-Kamerlingh Onnes Laboratory, Leiden University, P.O. Box 9504,
2300 RA Leiden, The Netherlands}
\author{K. Heeck}
\affiliation{Huygens-Kamerlingh Onnes Laboratory, Leiden University, P.O. Box 9504,
2300 RA Leiden, The Netherlands}
\author{W. Löffler}
\affiliation{Huygens-Kamerlingh Onnes Laboratory, Leiden University, P.O. Box 9504,
2300 RA Leiden, The Netherlands}
\begin{abstract}
Full characterization of surface acoustic wave (SAW) devices requires
imaging the spatial distribution of the acoustic field, which is not
possible with standard all-electrical measurements where an interdigital
transducer (IDT) is used as a detector. Here we present a fiber-based
scanning Michelson interferometer employing a strongly focused laser
beam as a probe. Combined with a heterodyne circuit, this setup enables
frequency- and spatially-resolved measurements of the amplitude and
phase of the SAW displacement. We demonstrate this by investigating
a 1 GHz SAW cavity, revealing the presence of frequency-overlapping
transverse modes, which are not resolved with an all-electrical measurement.
The frequency overlap of these transverse modes leads to mode superpositions,
which we analyze by quadrature decomposition of the complex acoustic
field.
\end{abstract}
\maketitle

\section{Introduction}

Surface acoustic waves (SAWs) are mechanical waves that travel along
the surface of a material and find many applications in modern technologies,
due to the ease of excitation on piezoelectric substrates via interdigital
transducers (IDTs \citep{joshiExcitationDetection1969,whiteDirectpiezoelectric1965,delsing2019surface2019}).
For instance, their surface confinement makes them useful in chemical
and biological sensing, where a compound placed on the free surface
of the piezoelectric material interacts with the propagating waves
\citep{paschkeFastSurface2017,chenUltrahighFrequencySurface2020,grateAcousticWave2000,langeSurfaceacoustic2008},
and manipulation of biological matter such as cells \citep{guoControllingcell2015,collinsTwodimensionalsinglecell2015,frankeSurfaceacoustic2010}.
The small acoustic wavelength at GHz frequencies allows for miniaturization
of electronic filters, finding applications in telecommunications
\citep{takaiHighPerformanceSAW2017,hashimotoSurfaceAcoustic2000a,yangAdvancedRF2023}.
Due to their long coherence times, and to the ability of interacting
with many different quantum systems, SAWs are used also in quantum
physics research where they can couple to a variety of two-level systems,
such as superconducting qubits \citep{manentiCircuitquantum2017,gustafssonPropagatingphonons2014,mooresCavityQuantum2018a,satzingerQuantumcontrol2018},
NV centers and quantum dots \citep{weibInterfacingquantum2018,schuetzUniversalQuantum2015,decrescentLargeSinglePhonon2022,metcalfeResolvedSideband2010b,satoTwoelectrondouble2017}.
Many of these applications require spatial confinement of the SAWs,
and this can be achieved by patterning periodic gratings (Bragg mirrors)
on the surface of the host material \citep{manentiSurfaceacoustic2016,takasuSurfaceacousticwaveresonators2019b,luschmannSurfaceacoustic2023b,msallFocusingSurfaceAcousticWave2020c},
to obtain an acoustic cavity. A typical configuration is a 1-port
resonator, consisting of two SAW mirrors enclosing an IDT, which can
be simultaneously used for SAW excitation and detection. 

The characterization of SAW devices is typically done in the frequency
domain with an all-electrical measurement using a Vector Network analyzer
(VNA) which, in the case of a SAW cavity, consists in measuring the
acoustic resonance spectrum \citep{camaraVectornetwork2017}. This
approach does not provide spatial information, which requires imaging
the acoustic field distribution in the SAW device, and can be done
with different techniques employing an external probe, such as atomic
force microscopy (AFM) \citep{hellemannDeterminingAmplitudes2022},
X-ray diffraction \citep{goddardStroboscopicsynchrotronXradiation1983,nicolasTimeresolvedcoherent2014a,hankeScanningXRay2023b},
and optical probes \citep{gualtieriLargearearealtime1996a,hisatomiQuantitativeoptical2023a,iwasakiTemporaloffsetdualcomb2022a,kamizumaHighSpeedLaser2005a,tagaOpticalpolarimetric2021a,rummelImagingsurface2021b}.
Among these, interferometric optical probes stand out for their simple
implementation and high sensitivity in displacement measurements \citep{knuuttilaScanningMichelson2000,kokkonenScanningheterodyne2008b,gollwitzerInterferometricObservation2006a,hashimotolaserprobe2011a,takahashiDevelopmenthighspeed2021a}.

Here we present a fiber-based scanning Michelson interferometer with
a heterodyne circuit, combining frequency- and spatially-resolved
measurements of the amplitude and phase of the SAW out-of-plane displacement.
We demonstrate this by investigating a planar-mirror SAW cavity device,
operating at 1 GHz. First we measure the acoustic resonance spectrum
at different spatial positions in the SAW cavity, which reveals the
presence of frequency-overlapping transverse modes. We compare this
measurements with an all-electrical VNA measurement, where the transverse
modes can not be resolved. Then we image the transverse modes by acquiring
spatial maps of the acoustic field distribution in the SAW resonator.
The frequency overlap between transverse modes leads to mode superposition,
which can result in unconventional mode profiles. We investigate these
effects by quadrature decomposition of the complex acoustic field
maps.

\section{Experimental Setup}

\begin{figure*}[t]
\includegraphics[width=1\linewidth]{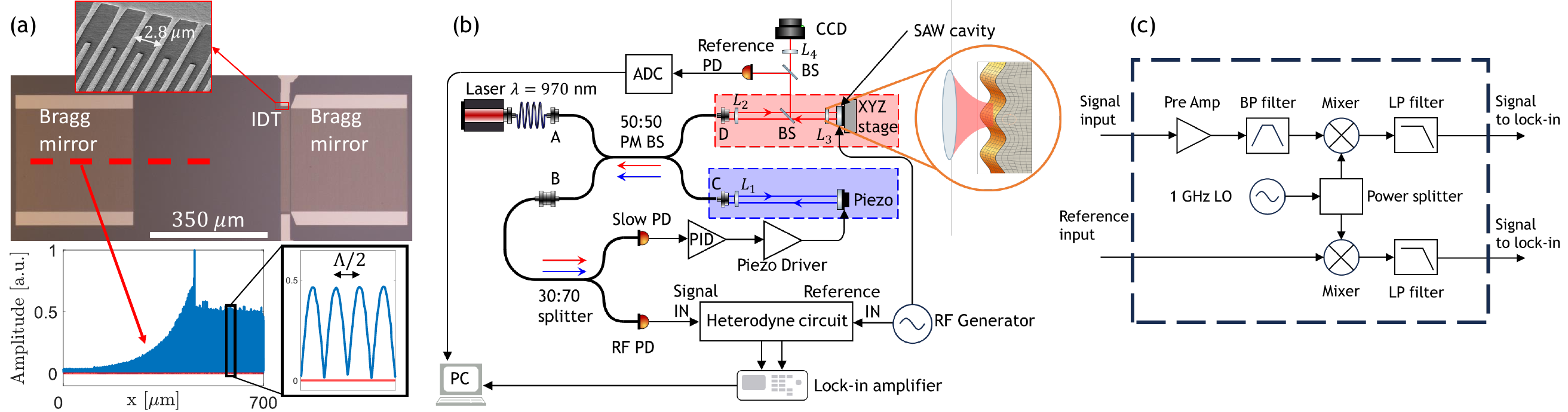}\caption{\protect\label{fig:Setup_panel}(a) Optical microscope image of the
SAW cavity device, and an electron microscope picture of the IDT metal
fingers (top inset). In the bottom panel we show an example of the
displacement field, measured with the optical interferometer at the
position indicated by the red dashed line. The exponential decay of
the field inside the Bragg mirrors is visible, as well as the nodes
and antinodes of the standing waves (bottom right). (b) Schematic
of the fiber-based scanning optical interferometer. (c) Schematic
of the custom-built $1\,\mathrm{GHz}$ heterodyne circuit used to
convert the $~1\,\mathrm{GHz}$ SAW signal from the RF photodiode
into an intermediate-frequency signal in the range of $1-48\,\mathrm{MHz}$.}
\end{figure*}

\subsection{SAW device}

The device investigated in this paper is a planar SAW cavity, as shown
in Fig. \ref{fig:Setup_panel} (a). This device has been nanofabricated
via e-beam lithography and e-beam evaporation of aluminum on a (001)-cut
GaAs substrate. It consists of an IDT with 10 metal finger pairs,
placed inside a cavity formed by two planar mirrors, each one made
of 250 metal fingers. The metal thickness of the IDT and the mirrors
is $h=50\,\textrm{nm}$, the centre-to-centre finger spacing is $p=1.4\,\mathrm{\mu m}$
and the finger width is $d=700\,\textrm{nm}$. The two mirrors are
placed at a distance of $L_{\mathrm{cav}}=470\,\mathrm{\mu m}$ from
each other, and the length of the fingers on the transverse direction
is $L_{y}=308\,\mathrm{\mu m}$. The device is oriented along the
{[}110{]} direction, for which the speed of the (Rayleigh) surface
acoustic wave is $v_{SAW}\simeq2860\,\mathrm{m/s}$ , resulting in
a wavelength $\Lambda\simeq2.8\,\mathrm{\mu m}$ at the frequency
$f_{SAW}=1.022\,\mathrm{GHz}$. 

\subsection{Fiber-based scanning Michelson interferometer}

The optical setup shown in Fig. \ref{fig:Setup_panel} (b) is a scanning
Michelson interferometer implemented with a polarization-maintaining
single-mode fiber coupler as the beam splitter (PM BS). Light from
a narrow-linewidth fiber-coupled laser (TOPTICA DL PRO) with $\lambda=980\,\mathrm{nm}$
enters the fiber coupler through port A, and is split into the sample
(port D) and reference arm (port C). In the reference arm, the light
is focused onto a mirror by a single aspheric lens $L_{1}$, and back
reflected into the fiber. In the sample arm, light from port D is
first collimated by an aspheric lens ($L_{2})$, and then strongly
focused onto the GaAs-based SAW device by an aspheric lens with 0.55
NA ($L_{3}$), resulting in a spot size of $2w_{0}=2.8\,\mathrm{\mu m}$,
where $w_{0}$ is the beam waist radius. 

Due to Fresnel reflection, at $\lambda=980\,\mathrm{\mu m}$, the
GaAs surface acts as a partial mirror with reflectivity $R\simeq0.3$,
and the reflected light is coupled back into the fiber coupler through
the same port D. The reflected light from the reference and sample
arm of the interferometer is then recombined through the same fiber
coupler, and exits through port B. After port B, the recombined light
is split again by a fiber splitter (30:70 splitting ratio). The weaker
part is sent to a slow photodiode (Thorlabs PDA36-EC), the other to
a radio frequency photodiode (New Focus 1514). The slow photodiode
(slow PD) is used to generate the error signal used for stabilization
of the interferometer (side-of-fringe lock). This signal is first
sent to a PID controller, and then to a piezo driver used to actuate
the piezo element on which the mirror in the reference arm of the
interferometer is attached. 

The GHz signal detected with the radio frequency photodiode (RF PD)
is down-converted to an intermediate frequency signal in the range
$1-48\,\mathrm{MHz}$ by the heterodyne circuit, and then measured
with a lock-in amplifier (Zurich Instruments HF2LI, $50\,\mathrm{MHz}$
bandwidth). The origin of the $1\,\mathrm{GHz}$ fluctuations in the
optical power lies in the interference between the light from the
reference arm, and the light from the sample arm of the interferometer,
which is phase-modulated at $1\,\mathrm{GHz}$ due to the surface
displacement associated to the SAWs. In the sample arm of the interferometer,
two pellicle beam splitters redirect part of the back reflected light
to a CCD, used to image and align the SAW device, and to a reference
photodiode (reference PD), which is used to record simultaneously
a reflectivity image of the SAW device to correlate the displacement
maps to the device structure. 

Finally, the SAW devices is mounted on a 3-axes nanopositioning stage,
where translation along the optical ($z$) axis allows to adjust the
focus of the laser beam, while the $x$ and $y$ axes allow to scan
the laser focus over the SAW device. The imaging of the SAW displacement
field is done line-by-line with continuous scanning of the $x$ axis
of the nanopositioner, simultaneously recording the lock-in signal
and the encoder signal of the $x$ translation stage, allowing fast
scans with a maximum resolution of $\sim7\,\mathrm{nm}$. An example
of such measurement is shown in the bottom panel of Fig. \ref{fig:Setup_panel}
(a), where we plot the amplitude of the standing SAWs over a line
in the middle of the transverse direction of the SAW cavity, resulting
in a spatial $x-\mathrm{periodicity}$ of half the acoustic wavelength
$\Lambda/2$, as expected for standing waves.

\subsection{Heterodyne circuit}

In order to perform heterodyne measurements, we need to compare the
signal at the output of the photodiode to a reference signal. For
this we use a power splitter to split the signal generated by the
RF generator shown in Fig. \ref{fig:Setup_panel} (b), one part is
used to drive the IDT in the SAW cavity at a frequency $f_{SAW}\sim1\,\mathrm{GHz}$,
and the other part is used as the reference signal in the heterodyne
circuit. In the heterodyne circuit shown in Fig. \ref{fig:Setup_panel}
(c), a local GHz oscillator is used to generate a signal at the frequency
$f_{LO}=f_{SAW}-f_{IF}$, where $f_{IF}$ is the intermediate frequency
at which we want to downconvert the RF photodiode signal. 

The signal generated by the local oscillator is split in two by a
power splitter in order to be mixed with the reference and input signal.
The reference signal is first mixed with the local oscillator, and
then low-pass filtered, obtaining at the output a reference signal
at the intermediate frequency $f_{IF}$. On the other hand, the low-amplitude
input signal is pre-amplified, band-pass filtered, mixed with the
local oscillator and finally low-pass filtered, obtaining at the output
a signal at the intermediate frequency $f_{IF}$. 

These two outputs are then fed into the lock-in amplifier: the output
from the reference port is used as an external oscillator in the lock-in
amplifier, to demodulate the output coming from the signal port. As
a result, the lock-in amplifier can measure amplitude and phase of
the SAW displacement at the intermediate frequency $f_{IF}$. 

When we want to change the drive frequency of the IDT $f_{SAW}$,
we change accordingly also the frequency of the local oscillator $f_{LO}$,
such that the intermediate frequency is kept constant at the arbitrary
value of $f_{IF}=22\,\mathrm{MHz}$. Due to the bandwidth of the band-pass
filter, $f_{SAW}$ is limited to $900-1370\,\mathrm{MHz}$.

\subsection{Voltage-displacement calibration}

The voltage signal measured by the lock-in amplifier is converted
into the displacement associated to the SAWs by considering the overall
gain of the interferometer:

\begin{equation}
\zeta\ =\ \frac{\lambda}{4\pi P_{half}}\frac{1}{C_{RF}\times G_{LI}\times r}\times10^{6}\left[\frac{\mathrm{pm}}{\mathrm{\mu V}}\right],\label{eq:calibration_displacement}
\end{equation}
where the first term $4\pi P_{half}/\lambda$ gives the change in
the optical power caused by a given flat-surface displacement for
an interferometer locked at the side of the fringe. $\lambda$ is
the optical wavelength, $P_{half}=170\,\mathrm{\mu W}$ is the optical
power at which we side-of-the-fringe lock the interferometer, measured
at the RF photodiode, $C_{RF}=700\,\mathrm{V/W}$ is the conversion
factor between input optical power and voltage at the output of the
RF photodiode, $G_{LI}=20.6$ is the total voltage gain provided by
the heterodyne circuit, and $r=r(w_{0},\Lambda)$ is a reduction factor
caused by the finite size of the beam waist radius $w_{0}$ compared
to the acoustic wavelength $\Lambda$. 

As discussed in Appendix \ref{sec:reduction_factor}, the analytical
calculation of the reduction factor leads to

\begin{equation}
r(w_{0},\Lambda)=\exp\left(-\frac{\pi^{2}w_{0}^{2}}{2\Lambda^{2}}\right),\label{eq:reduction_factor}
\end{equation}
showing an exponential dependence on $w_{0}^{2}$. An accurate measurement
of the beam waist is therefore essential for a proper calibration,
and we measure it by using a variation of the well known knife-edge
technique \citep{arnaudTechniqueFast1971,skinnerMeasurementradius1972},
where the focused Gaussian beam is scanned over the boundary between
free GaAs surface, and the aluminium contact pad of the IDT. 

Due to the two different reflectivities of aluminium and GaAs, in
reflection we can measure a transition over this region in the reflected
optical power recorded by the reference photodiode, resulting in a
measured beam waist radius $w_{0}=1.4\,\mathrm{\mu m},$ as shown
in appendix \ref{sec:Beam-spot-measurements}. In our experiment,
the SAW wavelength is $\Lambda=2.8\,\mathrm{\mu m}$, resulting in
a reduction factor $r=0.29$, and a total gain of the interferometer
equal to $\zeta=0.154\,\mathrm{pm/\mu V}$ rms.

\section{Transverse modes in a planar SAW cavity}

\begin{figure}[t]
\includegraphics[width=1\columnwidth]{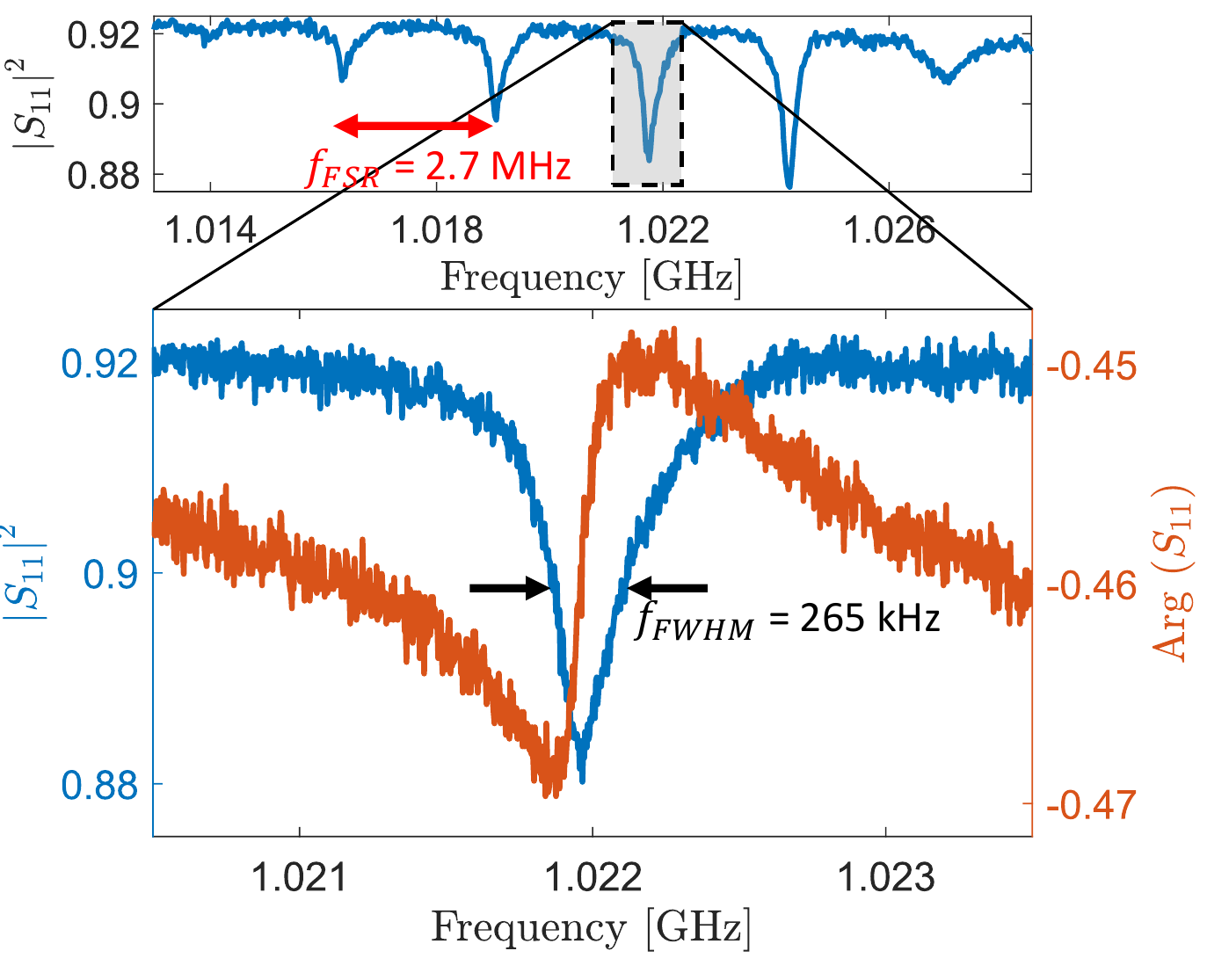}\caption{\protect\label{fig:VNA_measurement}Microwave reflectance spectrum
of the SAW cavity ($|S_{11}|^{2}$), measured with a VNA. The spectrum
shows multiple longitudinal modes separated by a free spectral range
$f_{FSR}=2.7\,\mathrm{MHz}$. In the bottom panel we show the cavity
resonance that is investigated with the optical setup, where the red
curve corresponds to the phase of the $S_{11}$ parameter. The measured
resonance frequency is $f_{c}^{VNA}=1.02197\,\mathrm{GHz}$, and the
full width half maximum $f_{FWHM}=265\,\mathrm{kHz}$.}
\end{figure}

\begin{figure}[t]
\includegraphics[width=1\columnwidth]{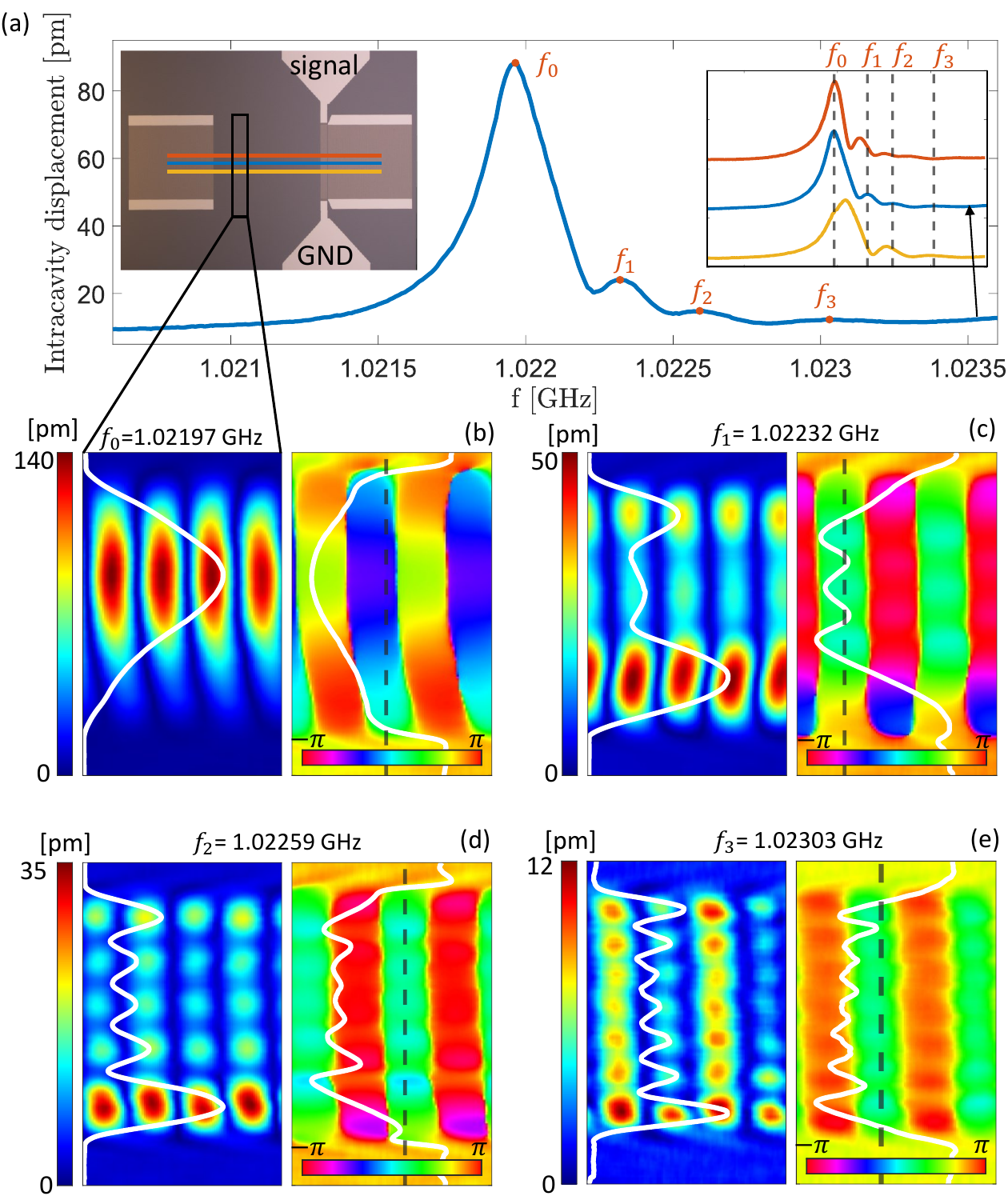}

\caption{\protect\label{fig:transverse_modes_and_freq_spectrum}Optical measurements
of the transverse modes. (a) Intracavity displacement as a function
of frequency, measured in the middle of the transverse direction of
the cavity. In the right inset we show three resonance spectra measured
at the different transverse positions indicated in the left inset,
demonstrating significant peak shifting and broadening. (b) - (e):
amplitude (left panels) and phase (right panels) of the transverse
modes, measured at the indicated peak frequencies, in the area indicated
by the black box in (a). The different transverse modes are axially
symmetric, showing respectively 1, 3, 5 and 7 lobes on the transverse
direction. The white lines show the cross section of the mode profile
measured at the position of an antinode as indicated by the dashed
lines in the phase plots.}
\end{figure}
Here we investigate the planar SAW Fabry-Perot cavity shown in Fig.
\ref{fig:Setup_panel} (a). First we characterize the cavity by means
of an all-electrical measurement done with a vector network analyzer
(VNA), then with the interferometric optical setup. By comparison,
we show how the latter approach allows for a better understanding
of the cavity modes and a better estimation of the cavity parameters.
The all-electrical VNA measurement is performed by driving the IDT
in the cavity with a vector network analyzer (NANO VNA), and measuring
the reflected electrical power. At a resonance of the cavity there
is conversion between electrical energy and mechanical energy, therefore
we expect to observe a dip in the measured reflected power. By performing
this measurement at different frequencies, we obtain the microwave
reflectance spectrum of the SAW cavity ($|S_{11}|^{2}),$ showed in
Fig. \ref{fig:VNA_measurement}. 

The device shows several resonance frequencies around 1 GHz, with
a free spectral range $\Delta f_{FSR}\ =2.7\,\mathrm{MHz}$, corresponding
to an effective cavity length $L_{eff}=530\,\mathrm{\mu m}$. By comparison
with the geometrical cavity length $L_{cav}=470\,\mathrm{\mu m}$,
we obtain a penetration depth into the Bragg mirror of $L_{p}\simeq30\,\mathrm{\mu m}$.
In the rest of this chapter, we will focus on the third reflection
dip, which has a resonance frequency measured by the VNA of $f_{c}^{VNA}\,=\,1.02197\,\mathrm{GHz}$,
and shows an asymmetric behavior on the high-frequency side, visible
in the bottom panel in Fig. \ref{fig:VNA_measurement}. This asymmetry
in the reflection dip is attributed to the excitation of frequency-overlapping
transverse modes, but there is no clear structure corresponding to
the individual peaks of these modes. 

This is because the IDT can not distinguish the modes spatially, due
to its transverse spatial extension. In fact, at a particular excitation
frequency close to the resonance of the fundamental mode, we have
superposition of different cavity modes, each one with a different
transverse spatial profile, different amplitude and different phase.
This leads to a complex frequency- and spatially-dependent mode superposition,
and since the IDT is extended along the full transverse length of
the cavity $L_{y}=308\,\mathrm{\mu m}$, it measures a spatial average
of this mode superposition, which in our case does not show a multi-peak
structure.

In order to separate these transverse modes in the frequency domain,
we measure the SAW cavity with the scanning Michelson interferometer.
Since this setup uses a focused laser beam as a small localized probe,
we expect the optically measured frequency spectrum to be dependent
on the position at which it is measured. Due to placement of the IDT
centered to the SAW cavity axis, we expect only symmetrical transverse
modes (\citep{campbellModellingtransversemode1991}). We drive the
IDT with a RF power of 15 dBm, and spatially scan a few periods of
the standing waves across the longitudinal $z$ direction of the cavity,
recording the peak value of the measured displacement amplitude. The
procedure is repeated for different frequencies, allowing us to reconstruct
the frequency spectrum of the intracavity field in the middle of the
cavity, as shown in Fig. \ref{fig:transverse_modes_and_freq_spectrum}
(a). In this frequency spectrum we now recognize several distinct
peaks, which correspond to different transverse modes as we show now. 

For this, we use the Michelson interferometer to image the spatial
acoustic field distribution inside the SAW cavity for the four peak
frequencies $f_{0}-f_{3}$, and we calibrate the measured displacement
using Eq. \ref{eq:calibration_displacement}. In particular, we image
a region inside the SAW cavity of $5.6\,\mathrm{\mu m}$ in the longitudinal
direction, and $400\,\mathrm{\mu m}$ in the transverse direction.
This is shown in Fig. \ref{fig:transverse_modes_and_freq_spectrum}
(b) - (e), where we plot both amplitude (left) and phase (right),
and we show the corresponding cross sections with a white curve. For
the amplitude maps, the cross section is taken at the position of
an antinode, while for the phase maps the cross section is taken at
the longitudinal position indicated by the black dashed line. 

These maps show transverse field profiles with an odd number of lobes,
in our case 1, 3, 5 and 7, corresponding to axially-symmetric transverse
modes. Moreover, by looking at the amplitude distribution of the acoustic
field, it is clear that there is an asymmetry with respect to the
center of the transverse direction: not only the field distribution
is shifted towards the upper side of the cavity, which is visible
especially in (b), but for multi-lobe modes, the lobe on the bottom
side of the cavity has a higher displacement than the one on the top
side. We do not know the origin of this shift and asymmetry, but we
hypothesize that it is related to the placement of the signal and
ground ports of the IDT. 

As mentioned above, due to the presence of the transverse modes which
overlap both in frequency and in space, the optically measured acoustic
resonance spectrum depends significantly on the transverse position
at which it is measured. This is shown in the right inset of Fig.
\ref{fig:transverse_modes_and_freq_spectrum} (a), where we measure
the acoustic resonance spectrum at different transverse positions
indicated by the colored lines in Fig. \ref{fig:transverse_modes_and_freq_spectrum}
(a) (left inset). From this measurement it is clear that, due to mode
superposition, the resonance spectrum of the displacement shows peaks
with different frequency and shape, depending on the transverse position
at which is measured. This makes it very hard, if not impossible,
to determine accurately the exact resonance frequencies of the different
transverse modes, and therefore the acoustic resonance spectrum shown
in Fig. \ref{fig:transverse_modes_and_freq_spectrum} (b) only gives
an estimate of these resonance frequencies. 
\begin{figure}[t]
\includegraphics[width=1\columnwidth]{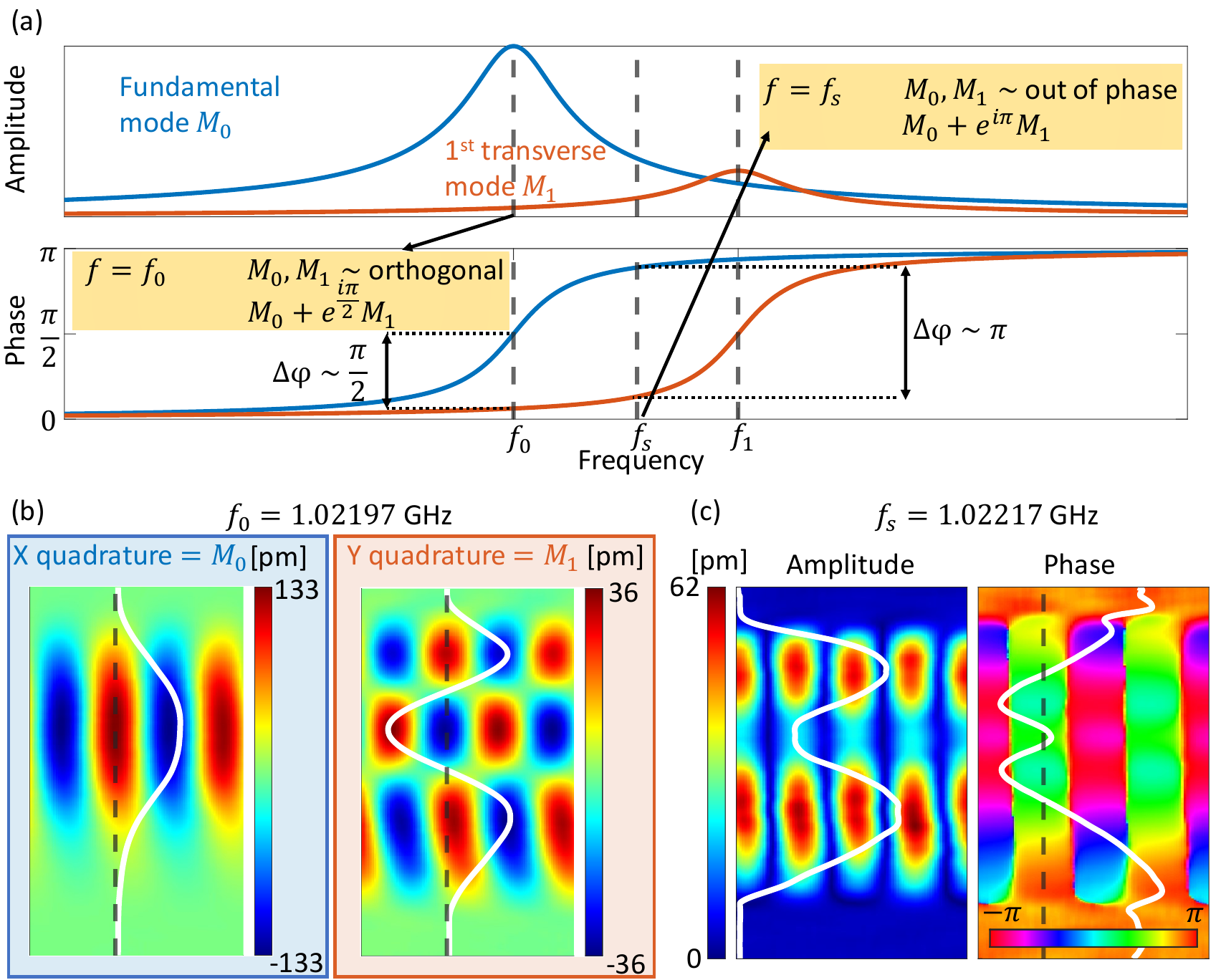}

\caption{\protect\label{fig:Modes_superposition}Mode superpositions. In (a)
we show the simulated amplitude and phase of two generic frequency-overlapping
cavity modes $M_{0}$ and $M_{1}$. At the resonance frequency $f_{0}$
of the fundamental mode $M_{0}$, the phase difference is $\Delta\varphi\simeq\pi/2$,
and the two modes are almost orthogonal, allowing mode decomposition
by quadrature analysis shown in (b). At the frequency $f_{s}$ in
between $f_{0}$ and $f_{1}$, the phase difference is $\Delta\varphi\simeq\pi$
and the two modes are almost out of phase with a comparable amplitude,
leading to destructive interference and a two-lobe structure (c).}

\end{figure}

Another consequence of the mode superposition is that the mode profiles
shown in Fig. \ref{fig:transverse_modes_and_freq_spectrum} (b) -
(e) do not correspond to transverse normal modes, but are a superposition
of multiple modes at the specific frequency at which they are measured.
To show this effect, and to show how we can retrieve the \textit{real}
mode profile, we consider the fundamental mode $M_{0}$, and the first
transverse mode $M_{1}$. Resonant modes in a cavity are enhanced,
with respect to the input field, and the ratio between the enhanced
field and the input field is referred to as susceptibility, described
by a complex Lorentzian \citep{aspelmeyerCavityoptomechanics2014b}
\begin{equation}
\chi(f)=\frac{1}{1-2i\frac{f-f_{c}}{\Delta f_{FWHM}}}.
\end{equation}
Here, $f_{c}$ is the mode frequency, and $\Delta f_{FWHM}$ its width.
In particular the susceptibility is characterized by a $\pi$ phase-shift,
as shown in Fig. \ref{fig:Modes_superposition} (a), where we plot
the simulated intracavity field enhancement for two generic cavity
modes. 

Let us now consider what happens when we optically measure the acoustic
field distribution at the frequency $f_{0}$. There, the amplitude
of the fundamental mode $M_{0}$ is maximum, while the amplitude of
the first transverse mode $M_{1}$ is strongly reduced due to the
detuning from its resonance frequency of $f_{1}$. The phase difference
between $M_{0}$ and $M_{1}$ is $\Delta\varphi\sim\pi/2$, so that
the modes are almost orthogonal. This means that in principle we should
be able to decompose the measured mode profile into the normal modes
$M_{0}$ and $M_{1}$, by analyzing the X and Y quadratures measured
with the lock-in amplifier. To do this, we take the amplitude and
phase of the acoustic field distribution measured at $f_{0}$ in Fig.
\ref{fig:transverse_modes_and_freq_spectrum} (b), we apply a phase
rotation of $\theta=1.15\,\mathrm{rad}$, needed to align the complex
acoustic field along the X and Y quadratures, and then we take the
real and imaginary part of this complex field, corresponding to the
X and Y quadratures. 

The result is shown in Fig. \ref{fig:Modes_superposition} (b), where
now we can better see the transverse profile of the $M_{0}$ and $M_{1}$
modes. Another interesting phenomenon caused by mode superposition
is the destructive interference of modes $M_{0}$ and $M_{1}$, visible
when we optically measure the acoustic field distribution at the frequency
$f_{s}=1.02217\,\mathrm{GHz}$. Since this frequency is in between
$f_{0}$ and $f_{1}$, the phase difference between $M_{0}$ and $M_{1}$
is now $\Delta\varphi\sim\pi$, causing destructive interference between
the two. The result is shown in Fig. \ref{fig:Modes_superposition}
(c), where the amplitude of the acoustic field distribution shows
a two-lobe structure. While in principle this could correspond to
an antisymmetric transverse mode, the symmetry observed in the phase
is proof that this is indeed a mode superposition between symmetric
modes, in our case $M_{0}$ and $M_{1}$.

Finally, as a demonstration of the accuracy of our measurements, we
estimate the surface displacement that we expect for the fundamental
mode of the planar SAW cavity. First we calculate the steady-state
phonon number $\bar{n}$ starting from the definition of cavity quality
factor that takes into account the stored energy:

\begin{equation}
Q=2\pi f_{0}\times\frac{E_{s}}{P_{loss}}.\label{eq:Q_factor}
\end{equation}
where $E_{s}$ is the energy stored in the cavity, and $P_{loss}$
is the power loss. At steady state, the stored energy is $E_{s}=\overline{n}\times hf_{0}$,
where $h$ is the Planck constant, and the power loss must be equal
to the input power in the cavity $P_{in}=\eta\times P_{\mu wave}$,
where $\eta$ is the fraction of power coupled to the cavity, and
$P_{\mu wave}$ is the RF power sent to the IDT. The quality factor
can also be described in term of linewidth of the resonator $Q=f_{0}/f_{FWHM}$,
where $f_{FWHM}$ is the width at half maximum of the cavity resonance.
By comparing this definition to Eq. \ref{eq:Q_factor}, we can obtain
the steady-state phonon number

\begin{equation}
\bar{n}=\frac{\eta\times P_{\mu wave}}{hf_{0}}\frac{1}{2\pi f_{FWHM}}.
\end{equation}
From the the microwave reflectance spectrum $|S_{11}|^{2}$, we estimate
$\eta=0.04$, and $f_{FWHM}=265\,\mathrm{kHz}$. The RF microwave
power sent to the IDT is $P_{\mu wave}=15\,\mathrm{dBm}$, obtaining
$\bar{n}=1.1\times10^{15}$. We can convert the steady-state phonon
number into a SAW displacement, by using the relation $u=u_{zpm}\times\sqrt{\bar{n}}$,
where $u_{zpm}$ is the zero-point motion, and for SAWs on GaAs it
has been numerically estimated to be $u_{zpm}\simeq1.9\,\mathrm{fm}/\sqrt{A[\mathrm{\mu m^{2}]}}$\citep{schuetzUniversalQuantum2015},
where $A$ is the surface area on which the mode is confined, expressed
in $\mathrm{\mu m^{2}}$. We estimate $A=L_{eff}\times L_{FWHM}\simeq9\times10^{4}\,\mathrm{\mu m^{2}}$,
where $L_{eff}=v_{SAW}/(2f_{FSR})=530\,\mathrm{\mu m}$ is the effective
cavity length corresponding to the free spectral range $f_{FSR}=2.7\,\mathrm{MHz}$
of our cavity as measured in Fig. \ref{fig:VNA_measurement}, and
$L_{FWHM}=170\,\mathrm{\mu m}$ is the FWHM of the SAW field profile
along the transverse direction in Fig. \ref{fig:transverse_modes_and_freq_spectrum}
(b). 

With these cavity parameters, we obtain an estimate of the SAW peak
displacement of $u\simeq210\,\mathrm{pm}$, which is of the same order
of magnitude as the optically measured value $u_{meas}=140\,\mathrm{pm}$
in Fig. \ref{fig:transverse_modes_and_freq_spectrum} (b). We note
that the estimated SAW peak displacement is an overestimation of the
real value, since we assumed that all the input power $P_{in}$ dissipated
in the IDT goes into excitation of the acoustic waves, whereas in
reality there are also Ohmic losses due to the finite resistance of
the IDT, metal bus bars and wirebonds.

\section{Conclusions and outlook}

We presented a fiber-based Michelson interferometer which, combined
with a heterodyne circuit, enables frequency- and spatially-resolved
measurements of amplitude and phase of surface acoustic wave displacements.
We used this setup to investigate transverse modes in a 1 GHz SAW
cavity, revealing the presence of higher order transverse modes which
were not resolved in all-electrical measurements. First, by imaging
the acoustic fields, we show that the transverse modes overlap in
frequency and space, leading to mode superpositions. This modifies
both the acoustic resonance spectrum of the cavity and the spatial
profile of the modes. Second, we show that the mode superposition
can be decomposed into normal modes by quadrature analysis. The combination
of frequency measurements and spatial imaging enables a more complete
characterization of SAW cavities, and of SAWs devices in general,
and can play a crucial role in optimization of SAW devices.

\section{Acknowledgements}

We would like to thank H. Visser for helping with the Heterodyne circuit
and for useful discussions. We acknowledge funding from NWO/OCW (QUAKE,
680.92.18.04; Quantum Software Consortium, No. 024.003.037), from
the Dutch Ministry of Economic Affairs (Quantum Delta NL), and from
the European Union’s Horizon 2020 research and innovation program
under Grant Agreement No. 862035 (QLUSTER).

\appendix

\section{Effects of the beam spot size on the measured interferometric signal}

\label{sec:reduction_factor}
\begin{figure}[!tph]
\includegraphics[width=1\columnwidth]{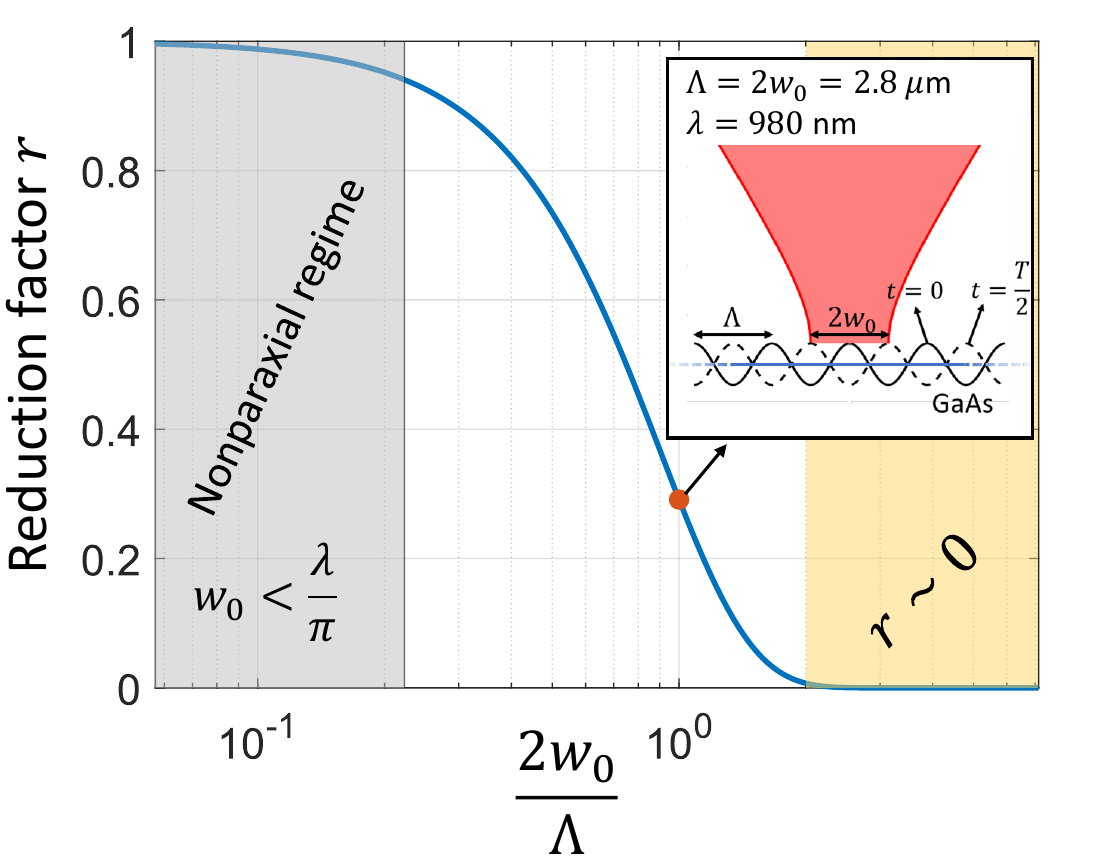}

\caption{\protect\label{fig:=000020SAW_GaAs_BeamSpotSize}Reduction factor.
The blue curve is the reduction factor from Eq. \ref{eq:reduction_factor},
as a function of the ratio between laser beam spot $2w_{0}$, and
the acoustic wavelength $\Lambda$ of SAWs. The red dot indicates
our experimental conditions with $\Lambda=2w_{0}=2.8\,\mathrm{\mu m}$,
and optical wavelength $\lambda=980\,\mathrm{nm}$. The gray shaded
area on the left with $w_{0}<\lambda/\pi$ corresponds to the nonparaxial
regime, where our calculation of the reduction factor is not valid.
The yellow shaded area on the right with $2w_{0}/\Lambda>2$, corresponds
to a region where $r\sim0$. In the right panel we show a sketch of
the experiment. We show the surface profile at time $t=0$ and $t=T/2$,
corresponding to half the oscillation period of the standing waves.}
\end{figure}
We discuss here the effects of the optical beam spot size on the measured
interferometric signal, for which we need to keep in mind the optical
interferometer sketched in Fig. \ref{fig:Setup_panel} (b). The gain
of the interferometer provided in Eq. \ref{eq:calibration_displacement}
has been calculated by taking into account the change in the optical
power at the photodiode, caused by the displacement of a reflective
flat surface in the sample arm of the interferometer. The effect of
the beam spot size is incorporated into the reduction factor $r(w_{0},\Lambda)$.

We can calculate this reduction factor by comparing the fluctuations
in the optical power at the photodiode, caused by reflection from
an oscillating flat surface ($P_{flat}^{\Omega})$, to the fluctuations
caused by reflection from the oscillating SAW surface ($P_{SAW}^{\Omega}$),
where $\Omega$ is the angular frequency of oscillation. Since the
heterodyne detection system detects the rms values of the optical
power fluctuations, we introduce $P_{flat}^{rms}$ and $P_{SAW}^{rms}$,
and define the reduction factor as

\begin{equation}
r=\frac{P_{SAW}^{rms}}{P_{flat}^{rms}},
\end{equation}
where $P_{SAW}^{rms}$ and $P_{flat}^{rms}$ denote the rms value
of $P_{SAW}^{\Omega}$ and $P_{flat}^{\Omega}$ respectively.

We now proceed to calculate these two fluctuations in the optical
power: looking at Fig. \ref{fig:Setup_panel} (b), first we calculate
the electric field coupled back into the fiber both in the reference
and sample arm of the interferometer, where the field in the sample
arm is modulated in phase by the oscillating surface, then we let
the two fields interfere and we keep only the $\Omega$ component
in the optical power fluctuation. Starting from the reference arm,
since the interferometer is locked at half of the interference fringe,
we define a reference field with unitary amplitude and constant phase
$E_{ref}=\exp(i\pi/2)$. 

In the sample arm of the interferometer, we want to calculate the
field back-reflected into port D of the fiber coupler for the two
cases of SAW oscillations and oscillations of a flat surface. This
field is given by the overlap integral between the field supported
by the optical fiber, and the field reflected by the oscillating reflective
surface. Since optical propagation is unitary, we can calculate this
integral at any plane in the sample arm of the interferometer, and
for simplicity we choose the plane of the reflecting surface. Here
the overlap integral is calculated between the image of the field
supported by the fiber $E_{fiber}$ and the reflected fields $E_{flat}$
and $E_{SAW}$. In the interferometer we can achieve close-to-unity
coupling back to the fiber, therefore we can assume that the image
of $E_{fiber}$ at the reflecting surface has the same beam waist
of the focused laser beam, leading to $E_{fiber}=\exp[-(x^{2}+y^{2})/w_{0}^{2}]$.
The reflected fields can be expressed as a generic Gaussian beam with
an additional phase term picked upon reflection from the oscillating
surface: 

\begin{align}
E & =\frac{2}{\pi w_{0}^{2}}\exp\left(-\frac{x^{2}+y^{2}}{w^{2}(z)}\right)\times\\
\times & \exp\left[-i\left(2k(z+\Delta z)+k\frac{x^{2}+y^{2}}{2R(z)}-\psi(z)\right)\right]\nonumber \\
 & \simeq\frac{2}{\pi w_{0}^{2}}\exp\left(-\frac{x^{2}+y^{2}}{w_{0}^{2}}\right)\times\exp\left(-2ik\Delta z\right),\nonumber 
\end{align}
where we drop the constant propagating phase $z$, and the phase terms
corresponding to the Gouy phase $\psi(z)$ and to the radius of curvature
$R(z)$. This approximation is valid for small surface displacements
close to the focus. We can then write $E_{flat}=E_{fiber}\times\exp\left(-2ik\Delta z_{flat}\right)$
and $E_{SAW}=E_{fiber}\times\exp\left(-2ik\Delta z_{SAW}\right)$.
Let us first consider the flat case, for which the surface displacement
is given by $\Delta z_{flat}=A_{0}\cos(\Omega t)$, where $A_{0}$
is the peak surface displacement, and $\Omega$ is the angular frequency
of the surface oscillation. The field coupled back to port D of the
fiber coupler is given by:

\begin{align}
E_{flat}^{'} & =\iint E_{fiber}^{*}\times E_{flat}\,dx\,dy\\
 & =\frac{2}{\pi w_{0}^{2}}\iint\exp\left(-2\frac{x^{2}+y^{2}}{w_{0}^{2}}\right)\times\nonumber \\
\times & \exp[-2ik\Delta z_{flat}]\,dx\,dy\nonumber \\
 & =\exp(-2ik\Delta z_{flat}).\nonumber 
\end{align}

The total power at the photodiode is given by the interference of
$E_{flat}^{'}$ with $E_{ref}$: 

\begin{align}
P_{flat} & =|E_{flat}^{'}+E_{ref}|^{2}\label{eq:P_flat}\\
 & =|\exp\left(i\frac{\pi}{2}\right)+\exp\left(-2ikA_{0}\cos(\Omega t)\right)|^{2}\nonumber \\
 & \simeq2+4kA_{0}\cos(\Omega t)+4k^{2}A_{0}^{2}\cos^{2}(\Omega t)\nonumber 
\end{align}
where we expanded for small displacements $A_{0}$. Keeping in mind
that the rms value of a time signal is defined as:

\begin{equation}
P_{rms}=\sqrt{\frac{1}{T}\int_{0}^{T}P^{2}(t)\,dt},
\end{equation}
we obtain the rms value of the $\Omega$ component $P_{flat}^{rms}=4kA_{0}/\sqrt{2}$.

We now repeat the same calculations for $E_{SAW}$, i.e. not assuming
a flat reflecting surface. In this case, the position-dependent surface
displacement is given by 

\begin{align}
\Delta z_{SAW}\  & =A_{C}\cos(\Omega t)\cos\left[K(x-x_{0})\right]\\
 & +\,A_{S}\sin(\Omega t)\sin\left[K(x-x_{0})\right],\nonumber 
\end{align}
which is the sum of two time-shifted standing waves, with peak displacement
$A_{C}$ and $A_{S}$ respectively. The surface displacement written
in this form is a superposition of traveling and standing waves along
the x direction, depending on the amplitudes $A_{C}$ and $A_{S}$.
$K\,=\,2\pi/\Lambda$ is the SAW wave number, $\Omega$ the angular
frequency, and $x_{0}$ is the transverse position of the laser beam
on the SAW device. We now carry out the calculations by considering
this generic displacement, and in the end we can separate between
the case of purely traveling waves by imposing $A_{C}\,=A_{S}\,=A_{0}$,
and purely standing waves by imposing $A_{S}\,=\,0$ and $A_{C}\,=\,A_{0}$.
In this way, the standing waves are defined such that at position
$x_{0}\,=\,0$, the laser beam is focused on an antinode, and where
$A_{0}$ is the peak surface displacement. The field that couples
back to the fiber is now given by
\begin{align}
E_{SAW}^{'} & =\iint E_{fiber}^{*}\times E_{SAW}\,dx\,dy\\
 & =\iint E_{fiber}^{2}\times\exp(-2ik\Delta z_{SAW}).\nonumber 
\end{align}

Due to the small SAW displacement, we can expand the exponential containing
$\Delta z_{SAW}$, and we can ignore all terms containing $\sin(Kx)$
since they are even functions and their integration is zero. The overlap
integral therefore becomes
\begin{align}
E_{SAW}^{'}\, & =\frac{2}{\pi w_{0}^{2}}\,\int\exp\left(-\frac{2y^{2}}{w_{0}^{2}}\right)\,dy\,\times\\
 & \times\int\exp\left(-\frac{2x^{2}}{w_{0}^{2}}\right)\ \left[1-2ik\gamma\cos(Kx)\right]\:dx,\nonumber 
\end{align}
where $\gamma\,=A_{c}\cos(\Omega t)\cos(Kx_{0})-A_{S}\sin(\Omega t)\sin(Kx_{0})$.
The integral can be solved by expanding the cosine term $\cos(Kx)=[\exp(iKx)+\exp(-iKx)]/2$,
and using the standard integral:
\begin{equation}
\int_{-\infty}^{+\infty}\exp(-ax^{2}+ibx)\,dx\,=\,\sqrt{\frac{\pi}{a}}\exp\left(-\frac{b^{2}}{4a}\right).
\end{equation}

The result is given by:

\begin{equation}
E_{SAW}^{'}\simeq1+2ik\gamma e^{-B},
\end{equation}
where $B=-K^{2}w_{0}^{2}/8$. Similarly to what we did in Eq. \ref{eq:P_flat},
we calculate the total power at the photodiode, resulting in the interference
of $E_{SAW}^{'}$ and $E_{ref}$: 
\begin{equation}
P_{SAW}=2-4\gamma e^{-B}+4k^{2}\gamma^{2}e^{-2B}.
\end{equation}

The $\Omega$ component of the optical power at the photodiode for
a SAW displacement is given by
\begin{equation}
P_{SAW}^{\Omega}\,=\,4k\gamma e^{-B}.
\end{equation}

Now, before we calculate the rms value, we need to separate the case
of traveling and standing waves: for traveling waves we impose $A_{C}=A_{S}=A_{0}$,
which results in $\gamma=A_{0}[\cos(\Omega t)\cos(Kx_{0})\text{-\ensuremath{\sin}(\ensuremath{\Omega}t)\ensuremath{\sin}(K\ensuremath{x_{0}})]}$.
We then obtain $P_{tSAW}^{rms}\,=4kA_{0}\exp(-B)/\sqrt{2}$, where
the subscript $t$ stands for traveling. For the case of standing
SAW waves, we impose $A_{S}=0$ and $A_{C}=A_{0}$, obtaining $\gamma=A_{0}\cos(\Omega t)\cos(Kx_{0})$,
and the rms change in the optical power is given by $P_{sSAW}^{rms}\ =\ P_{tSAW}^{rms}\times\left|\cos(Kx_{0})\right|$,
where the subscript $s$ stands for standing wave. The measured rms
change in the $\Omega$ component of the optical power caused by traveling
SAWs is therefore equivalent to the one caused by standing SAWs, measured
when the laser is on an antinode ($x_{0}=0$). By comparing this result
with the rms change in the optical power caused by the displacement
of a flat surface, we obtain the wanted reduction factor $r$:

\begin{equation}
r\,=\,\frac{P_{SAW}^{rms}}{P_{flat}^{rms}}\,=\,\exp\left(-\frac{K^{2}w_{0}^{2}}{8}\right)
\end{equation}

In our case $\Lambda=2.8\,\mathrm{\mu m}$ and $w_{0}=1.4\:\mathrm{\mu m}$,
resulting in $r=0.29$, as shown in Fig. \ref{fig:=000020SAW_GaAs_BeamSpotSize}
where we plot the reduction factor and we show our experimental conditions.
The total gain of the interferometer in Eq. \ref{eq:calibration_displacement}
is therefore $\zeta=0.154\,\mathrm{pm/\mu V}$ rms.

\section{Beam spot size measurement}

\label{sec:Beam-spot-measurements}
\begin{figure}[t]
\includegraphics[width=1\columnwidth]{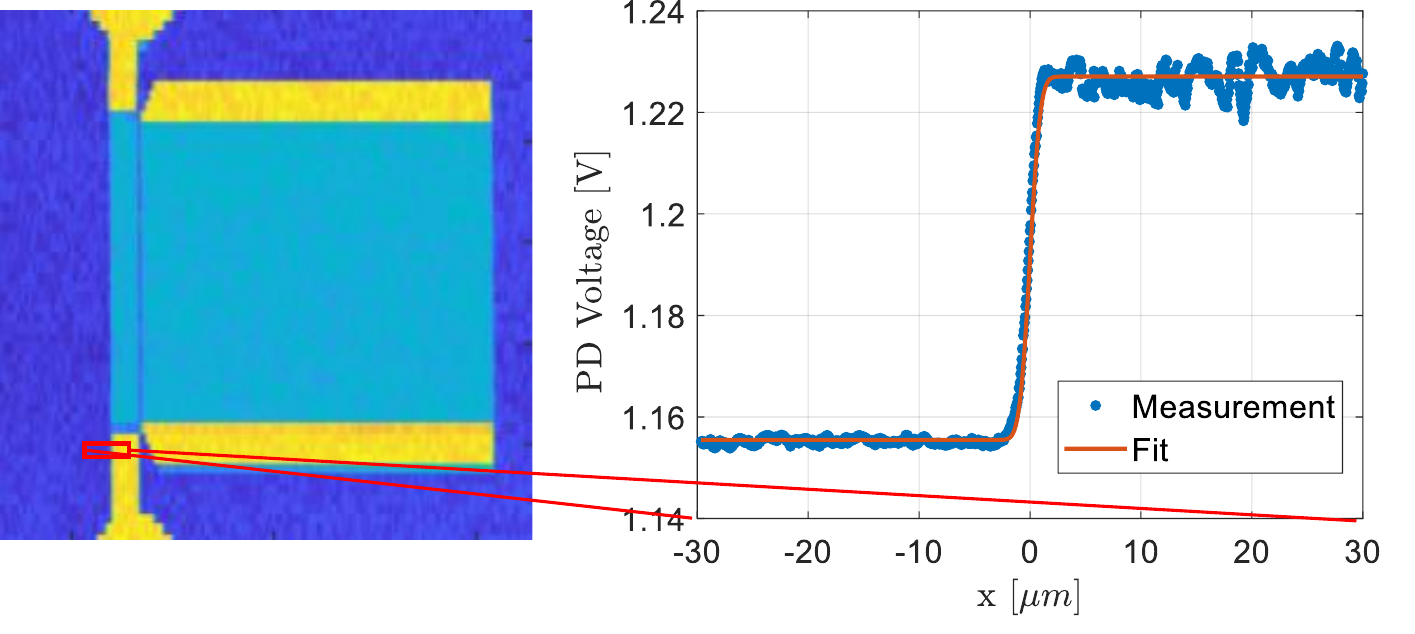}

\caption{Measurement of the beam spot size with a variation of the knife-edge
technique. Left panel: image of the IDT and Bragg mirror metal structure,
obtained with the reference photodiode in Fig. \ref{fig:Setup_panel}
(b). Right panel: photodiode voltage (blue) as a function of $x$
position, and fitted function (red), according to Eq. \ref{eq:fit_knife_edge}}
\label{fig:beamSpotSize}
\end{figure}

As discussed in the main text, the sine-like shape of the displacement
generated by a SAW affects the reduction factor $r=\exp(-K^{2}w_{0}^{2}/8)$.
Due to the exponential dependence on the beam waist radius, an accurate
measurement of $w_{0}$ is crucial for a proper calibration of the
measured voltage into a displacement. We perform this measurement
in-situ, using a variation of the knife-edge technique. Here we use
the boundary between GaAs substrate and the metal contact pad of the
SAW cavity IDT as a knife edge. We shine a focused laser beam on the
sample, we move the sample along the x direction, and using the reference
photodiode in Fig. \ref{fig:Setup_panel} (b), we measure the reflected
power over the boundary between GaAs and Al. An example of such measurement
is shown in Fig. \ref{fig:beamSpotSize} In a reference frame where
the boundary coincides with the origin of the x axis, the reflected
power is

\begin{equation}
P\ =\ P_{0}[r_{2}\ +\ r_{1}\ +(r_{2}-r_{1})\times erf(\frac{x\sqrt{2}}{w_{0}})],\label{eq:fit_knife_edge}
\end{equation}

where $P_{0}$ is the the optical power of the incident beam (in arbitrary
units), $r_{1}$ is the reflection coefficient for GaAs, $r_{2}$
is the reflection coefficient for Al, and $w_{0}$ is the beam waist
radius. In figure \ref{fig:beamSpotSize} is shown the region over
which such measurement has been performed, the measured data and the
fitting function. The fitted parameters are $w_{0}=1.4\,\mathrm{\mu m}$
and $r_{1}=0.30$. The expected reflection coefficient of GaAs at
$\lambda\simeq1\,\mathrm{\mu m}$ is $r_{GaAs}\ =\ $0.3097, which
is in very good agreement with the fitted $r_{1}$. \medskip{}

\rule[0.5ex]{1\columnwidth}{1pt}

\bibliographystyle{naturemagwV1allauthors}
\bibliography{bibliography}

\end{document}